\documentclass[12pt,a4paper]{article}
\usepackage[utf8]{inputenc}
\usepackage[a4paper,total={7in,9in}]{geometry}
\usepackage{graphicx}
\usepackage{float}
\usepackage{multirow}
\usepackage{amsmath}

\title{A feasibility study of a thorium fueled molten salt micro modular subcritical reactor using an electron accelerator}
\author{A Rummana
 \\ The University of Huddersfield, Huddersfield, UK \and  R J Barlow\\ The University of Huddersfield, Huddersfield, UK \and G Myneni \\ BSCE Systems, Inc., USA \and S M Saad \\ University of Technology and Applied Sciences - Ibra, Oman}
\date{December 2023}

\begin{document}

\maketitle

\def \U {{${}^{233}{\rm U}$}}
\def \Th {{${}^{232}{\rm Th}$}}
\def \Ta {{${}^{180}{\rm Ta}$}}
\def \Pb {{${}^{208}{\rm Pb}$}}
\def \Be {{${}^{9}{\rm Be}$}}
\def \FLiBe {{${\rm Li}_2{\rm BeF}_4$}}

\begin{abstract}
We present a design for a small subcritical molten salt thorium breeder reactor driven by an electron accelerator. Such a reactor could provide a safe and simple power source fuelled by thorium, without generating long-lived minor actinides. We use both Geant4 and MCNPX simulations to study the production of photons and photoneutrons, the criticality and the breeding in a simple conceptual design.
We show that the concept is on the edge of viability.
\end{abstract}

\section{Introduction}
Small Modular Reactors (SMR) are gaining interest globally \cite{SMR}, due to their many claimed advantages: 
\begin{itemize}
\item Flexible power generation, up to 300 MW/module.
\item Enhanced safety performance through advanced design.
\item Cost reduction through modularization, and used as single modules or in multiples.
\item In-factory fabrication that improves construction and scheduling cost
\item Deployment at  remote locations.
\end{itemize}


Micro modular reactors (MMRs) are a subcategory of SMR capable of generating up to 10 MWe. 
They are compact enough to be transported by truck, offering a flexible solution for energy production. As the demand for energy continues to rise, MMRs can be deployed incrementally, with each additional module providing additional power. Furthermore, MMRs can be seamlessly integrated with renewable energy sources to satisfy the growing need for clean energy. Some notable MMRs currently under development include Energy Well (Czech Republic), MoveluX (Japan), U-Battery (Urenco, UK) and AURORA (Oklo, US), Westinghouse eVinci (USA) and MMR (USA)~\cite{ARIS}. 

SMRs share the problem of long-term storage for radioactive waste. 
which demands secure disposal solutions spanning hundreds of thousands of years. 
However, use of thorium rather than uranium can lessen this requirement, reducing minor actinide production by approximately 100-fold compared to uranium-based fuels~\cite{Nifenecker, Asiya}. 
 and also has advantages of
proliferation resistance,
and greater natural abundance.
There are advantages for operating thorium systems
in subcritical mode,
driven by an accelerator whose current controls the power level~\cite{AccApp,Myrrha}. 

Molten salt reactors have advantages of easy online fuel processing, high neutron economy, negative temperature-criticality coefficients  and better fuel utilization~\cite{Zhao}, though
their engineering and chemistry do bring challenges. 

Combining these arguments, we are led to consider 
the design of a thorium-fuelled molten salt accelerator driven MMR, and that is the topic of this paper.

As a further distinctive feature we consider using an electron accelerator as the driver, rather than the more usual proton machine.  Proton accelerators are large and expensive, and inappropriate for s small reactor.
Electron accelerators are smaller and cheaper, though not nearly so effective at producing neutrons.  However neutron production can occur through a two stage process: 
the electrons produce an electromagnetic (EM) shower through Bremsstrahlung, and the gamma component produces neutrons off target nuclei, chiefly through the giant dipole resonance.


\subsection{Previous design history}

\begin{figure}[h!]
\centerline{\includegraphics[width=18 cm]{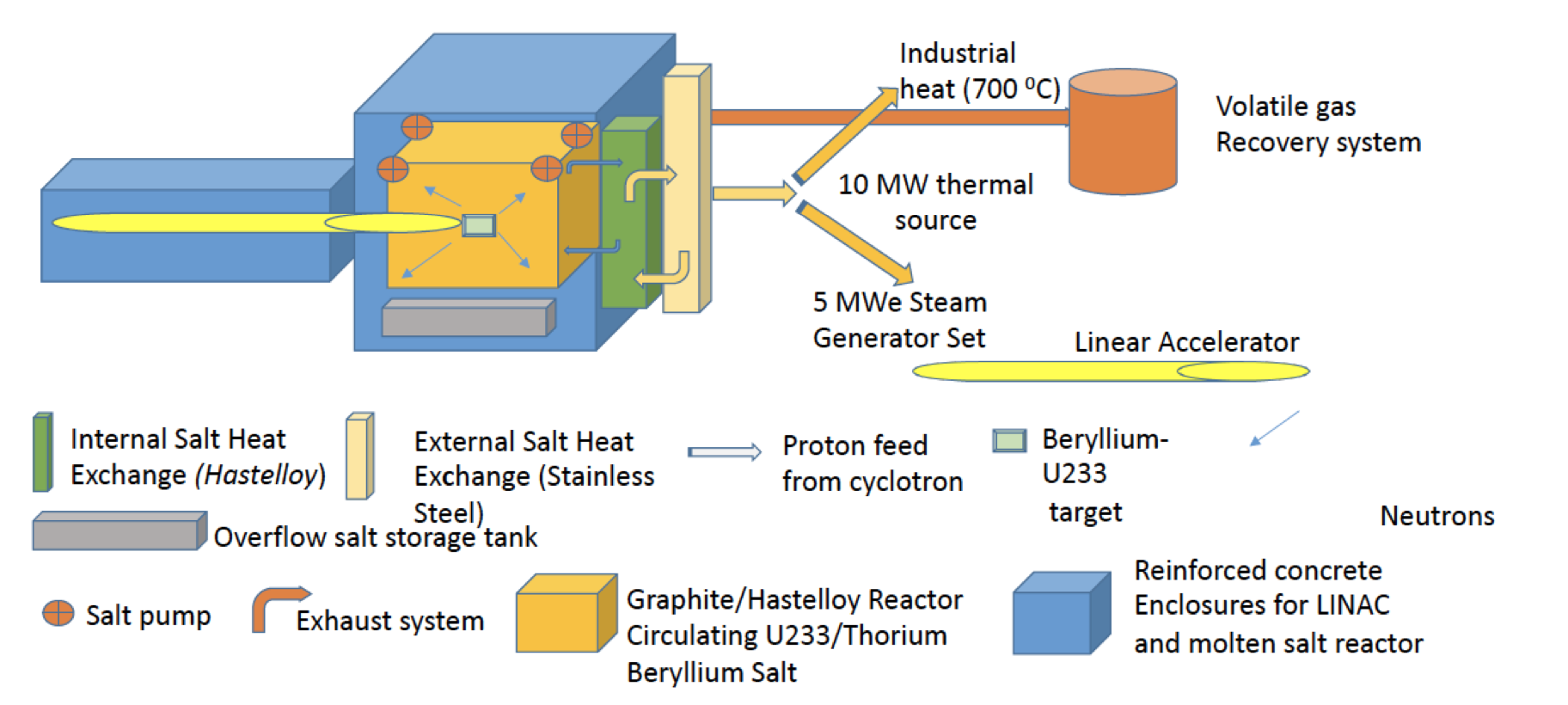}}
\caption{\label{fig:overview} An overview of the original design, taken from \cite{previous}. }
\end{figure}
This study is a development of the concept described previously~\cite{previous}
and shown in Figure~\ref{fig:overview}, though 
detailed neutronics calculations were not carried out. The new design, like the old, contains a linear accelerator and a molten salt system, specifically (FLIBE - \FLiBe ), containing dissolved fissile \U \ 
and fertile \Th. 
However in the new design the target is not separate from the salt, and it uses an electron rather than a proton LINAC.  

\subsection{Electron based neutron production}

Designs of electron based neutron sources have been presented previously~\cite{Abalin,Liu,Feizl,Zelinsky}. The feasibility of an electron based accelerator for nuclear waste transmutation is examined by Y.Liu~\cite{Liu} and is compared with proton based accelerator. While the advantages of electron based systems are reported as low radiation damage, wide availability at 100 MeV, low cost of the facility; the disadvantages are low efficiency of neutron generation rate, low transmutation efficiency, low thermal power and electricity generation as compared to proton based facilities. Among the various target materials considered (lead, tantalum, bismuth, thorium, uranium and tungsten) the highest neutron yield is observed for uranium.

A neutron source based on a sub critical assembly driven by an electron LINAC was developed at NSC KIPT~\cite{Zelinsky}. Studies have been done for a neutron source at Frascati using a 510 MeV electron beam~\cite{DAFNE1,DAFNE2}. They use a tungsten target, noting that with increasing Z the size of the giant dipole resonance increases, and the nucleon binding energy falls. They attain of the order of 0.2 neutrons per incident electron. 

A recent study by Gillespie et al \cite{Gillespie}
simulates the production of neutrons from a 10 MeV electron beam and a sandwich of foils of tantalum (to produce photons) and erbium deuteride, ${\rm ErD}_3$ to produce photoneutrons. Their best performance is $10^{-4}$ photoneutrons per electron. Interestingly, they find that the tantalum has no beneficial effect and their suggested design is for a pure ${\rm ErD}_3$ target.

\subsection{Comparison with similar designs}

India has recently commenced fundamental studies to conceptualize the design of the Indian Molten Salt Breeder Reactor (IMSBR). These studies encompass various design options and possibilities, with a focus on reactor physics and thermal hydraulic design~\cite{IMSBR}.

Most of the globally developed MMRs are high-temperature micro reactors featuring TRi-structural ISOtropic particle fuel (TRISO), primarily uranium-based~\cite{ARIS}. The concept of molten salt micro reactors employing thorium-based fuel represents a pioneering approach in the field of breeder reactors. India possesses a substantial reserve of thorium, yet despite the design of the 5 MWth Indian Molten Salt Breeder Reactor (IMSBR) at BARC~\cite{BARC}, there is currently no smaller scale prototype of a Micro Modular Reactor (MMR) capable of demonstrating feasibility for deployment.

This design is distinct from these with several unique features:
\begin{enumerate}
  \item Adoption of thorium-based fuel rather than the conventional use of uranium for a Micro Modular Reactor.
  \item Consideration of a notably lower power output (1 MWe).The choice of a much smaller power output, coupled with the utilization of thorium-based fuel, is anticipated to simplify the licensing and deployment processes.
  \item Using an electron accelerator rather than a proton machine.
  \item Using molten salt as the fuel, simplifying construction and unifying fuel, coolant, moderator and target.

\end{enumerate} 

The primary focus of this paper is optimizing photoneutron production, exploring the criticality and neutronics of fertile-to-fissile conversion, and demonstrating the feasibility of the proposed design. The overarching objective is to establish the reactor’s capability to achieve a viable and sustainable operation. It is very much a conceptual study: 
it is as yet too early for
detailed design considerations.

\section{Photon Production}

\begin{figure}[h]
\centerline{\includegraphics[width=16 cm]{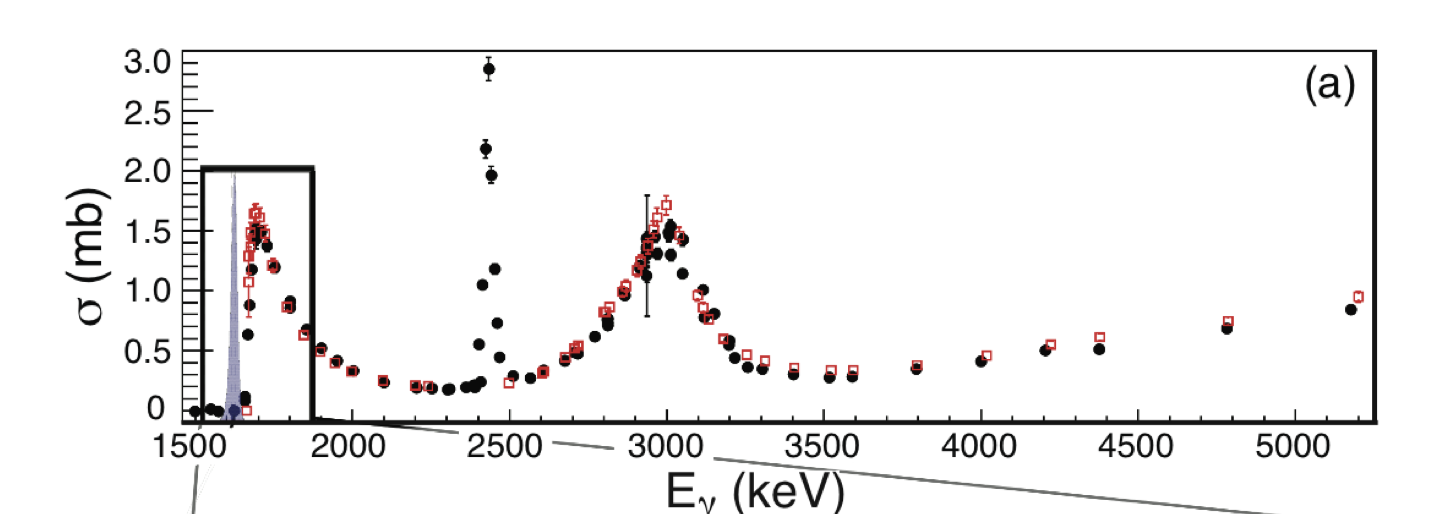}}
\caption{\label{fig:crosssection} The cross section for berylliuum break up as a function of photon energy. Taken from~\cite{xsect}.}
\end{figure}

Studies of EM showers in 
molten salt with various additives were  done using the EGS program~\cite{EGS} and Geant4~\cite{Geant4, AllisonNucl, AllisonIEEE}.
Using two different programs gives a check on the validity of our conclusions. The statistical precision of the simulation is 2\% to 6\% for Geant4. EGS, using 10,000 incident electrons, gives results with a statistical precision of order 1\%. The cross section for beryllium break-up, shown in Figure~\ref{fig:crosssection}, has a threshold
at 1.7 MeV. There is therefore no interest in
tracking the shower development of electrons and photons below this number.

An EGS geometry was arranged using a slab of FLiBe 30 cm thick. Geant4 geometry was constructed using a 30 cm long cylinder of radius 32.6 cm giving 100 litres of salt. The cylinder contains a mixture of FLiBe, uranium (\U) and thorium (\Th) whose concentrations are adjusted to maintain criticality slightly below 1 -- the exact figure is a crucial design choice later, at this stage all that is required is to compare equivalent mixtures.

Ideally one would optimise the photon spectrum to match the cross section. In practice there is no
scope for such fine tuning, and as a figure of merit we use the track length in the molten salt for photons about the threshold of 1.7 MeV.

The original design~\cite{previous} included small beads of tantalum in the molten salt, to act as sources of photons.
Calculations were done to study the photon production for different sizes and concentrations of tantalum beads.
Both the Geant4 and the EGS calculations pointed to the conclusion that the tantalum gave little benefit at best, and
in many cases degraded the performance.  With hindsight the reason is obvious: in the MeV electromagnetic $e/\gamma$ cascade one target electron in the fuel is very much like another, with small effects from the binding energy.
High $Z$ components can be useful if there is a need for the shower to develop over a short distance, but this is not the case here, with the reactor vessel dimensions larger than the shower size. It was concluded that adding tantalum beads served no purpose for producing the desired high energy photons.

\begin{figure}[H]
\centerline{\includegraphics[width=16 cm]{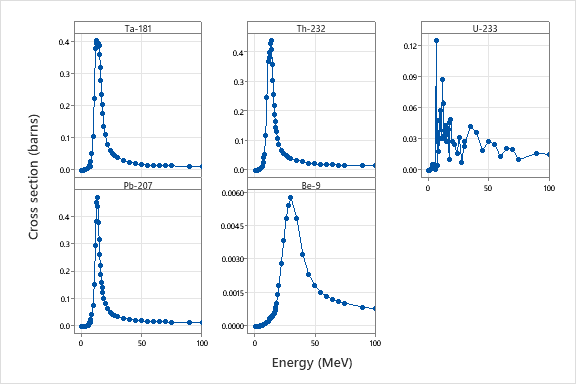}}
\caption{\label{fig:XS} Photonuclear cross section on tantalum, thorium, uranium, lead and beryllium  (Data taken from \cite{ENDF})
}
\end{figure}

However -- and again with hindsight this should have been anticipated -- 
it was found that adding tantalum to pure FLiBe did increase the yield of photoneutrons.
The size of the giant dipole resonance is proportional to $Z^2$, and the
photoneutron cross sections for high $Z$ isotopes are much greater than
that for beryllium, as shown in Figure~\ref{fig:XS}.  As the proposed fuel mixture contains appreciable amounts of \U\ and \Th, it may well be the case that these, rather than tantalum, provide
as much high-Z target material as can be useful. If further high Z material is required it would be better deployed as ${}^{208}{\rm Pb}$, as that does not absorb neutrons whereas
tantalum does, as discussed in section~\ref{Criticality}.

\subsection{Fuel composition}
In order to get optimum (that is maximum) number of neutrons per electron, different fuel compositions were considered. The cross sections for the photonuclear process on \Ta, \Pb, \Th, \Be\ and \U\ are shown in Figure~\ref{fig:XS}. It is clear from the figures that the cross sections of tantalum, lead and thorium are high for photo nuclear process around 10 MeV, whereas for uranium it is somewhat smaller, and the 
beryllium cross section is much smaller. So the high Z nuclei in the salt mixture give the neutrons. But they also have effects on the criticality, which will be considered in section~\ref{Criticality}.

To observe the effect from different elements, the following compositions are considered, each for one mole of FLiBe:
\begin{itemize}
\item U - 0.1 mole, Th - 1 mole
\item U - 0.15 mole, Th - 1.5 mole
\item U - 0.2 mole, Th - 2 mole
\item U - 0.1 mole, Th - 1 mole, Ta - 1 mole
\item U - 0.1 mole, Th - 1 mole, Pb - 1 mole
\end{itemize}

These high concentrations are driven, as will be seen, by the need for near-criticality, which drives up the \U \ concentration, and for
breedng replacement, which drives up the \Th\ concentration.  We assumed that the metals were added to the FLiBe salt; it may be that they would be incorporated in the form of flourides. The effect of any extra flourine was not considered in the calculations, but will not be significant.

\subsubsection{The Electron energy}\label{eEnergy}

The number of high energy photons in the shower increases with the electron energy, 
but so does the input power required. We quantify this by dividing the photon performance by the electron energy.
This dependence needs to be combined with the cost and affordability of electron LINAC in choosing the 
electron energy for the reactor.

\begin{figure}[H]
\centerline{\includegraphics[width=10 cm]{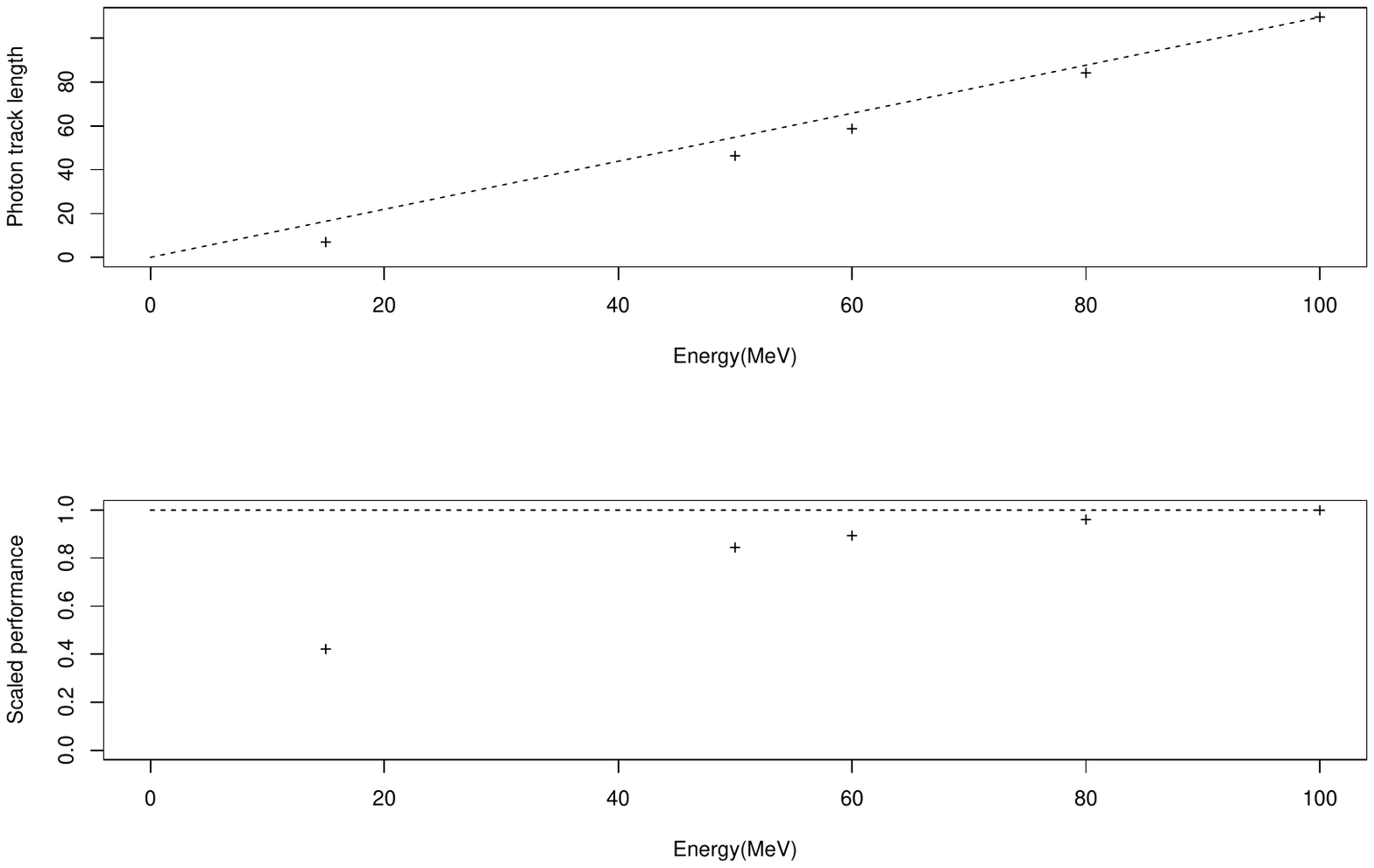}}
\caption{\label{fig:EGSEnergy} Photon track length and performance as a function of energy
(EGS)
}
\end{figure}
 An EGS study using effective track length  is shown in Figure~\ref{fig:EGSEnergy}.  The performance scales approximately linearly 
at high energies (the straight line is shown to guide the eye)
but falls off significantly at low energies.

\begin{figure}[H]
\centerline{\includegraphics[width=12 cm]{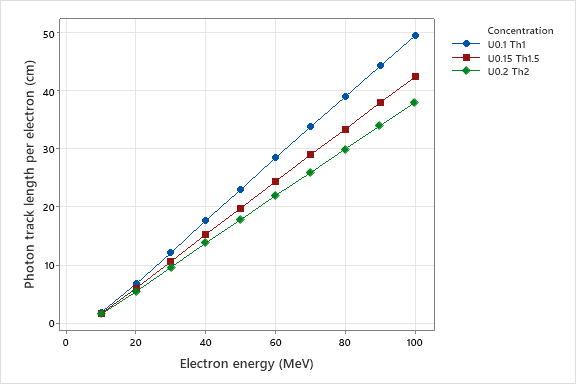}}
\caption{\label{fig:G4EnergyTL} Photon track length as a function of electron energy (Geant4)
}
\end{figure}

\begin{figure}[H]
\centerline{\includegraphics[width=12 cm]{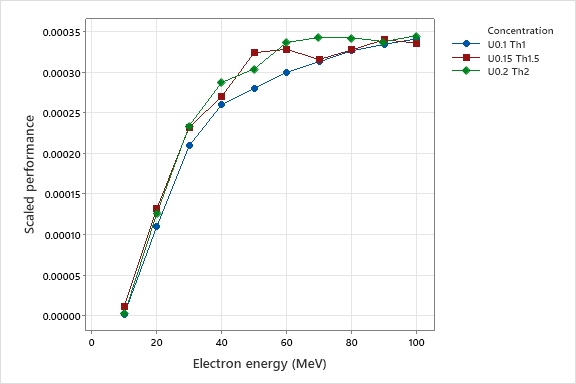}}
\caption{\label{fig:G4EnergySP} Photon performance as a function of electron energy (Geant4)
}
\end{figure}
The equivalent results from Geant4 are shown in Figure~\ref{fig:G4EnergyTL} and Figure~\ref{fig:G4EnergySP}, which also include different fuel composition. Figure~\ref{fig:G4EnergyTL} shows that the track length increases with the electron energy for all the chosen concentrations of uranium and thorium. However, increasing the concentration doesn't significantly effect the track length. So, in terms of track length 0.1 mole of Uranium and 1 mole of thorium is the optimum concentration. The scaled performance is shown in Figure~\ref{fig:G4EnergySP}. Above 60 MeV, the curves are almost flat while it decreases below it for all the chosen concentrations. So 60 MeV is taken as the ideal electron energy for photoneutron production in the present study.

\begin{figure}[H]
\centerline{\includegraphics[width=10 cm]{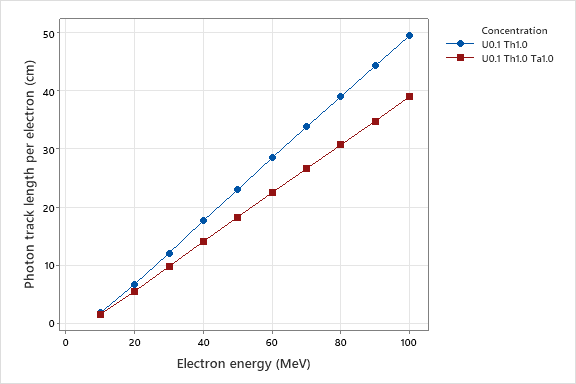}}
\caption{\label{fig:TLTa} Effect of adding tantalum on photon track length}
\end{figure}

\begin{figure}[H]
\centerline{\includegraphics[width=11 cm]{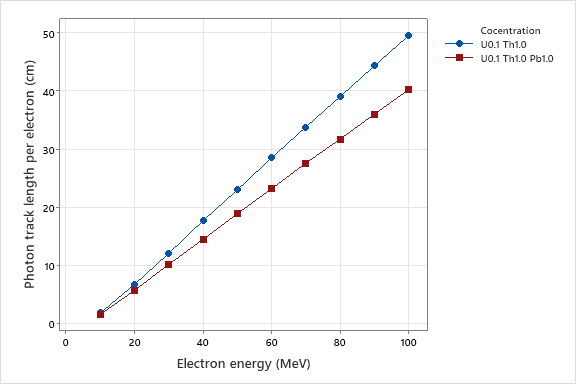}}
\caption{\label{fig:TLPb} Effect of adding lead on photon track length}
\end{figure}

Further Geant4 simulations were performed to see the effect of adding other elements. Since the photonuclear cross section for tantalum and lead are similar to that of thorium, these elements are then added with the same concentration as that of thorium. Figure~\ref{fig:TLTa} and Figure~\ref{fig:TLPb} show the effect of adding tantalum and lead in the mixture respectively for 0.1 mole of uranium and 1 mole of thorium. From these figures, it is clear that there is no benefit (in terms of track length) of adding them in the mixture.  

\section{Fissions}

In this preliminary study we consider only the neutronics of the salt and the added components, 
ignoring the effects of fission products and of intermediate isotopes such as protactinium.  This ideal could be achieved in principle by 
continuous processing of the circulating molten salt. In practice this will involve some challenging engineering and chemistry, and may not be completely achievable, but
any modelling would have to make specific assumptions about timescales for the power generation and for the reprocessing,
and these lie outside our scope. So we proceed, with the caveat that our results will represent an optimistic limiting behaviour. 

\subsection{Criticality of FLiBE/uranium fuel}

The criticality of the reactor was studied according to the well-established MCNPX~\cite{MCNP} procedure: we used 100 cycles, ignoring the first 10 in taking the average.
 
A cylindrical geometry was used, initially
with a radius of 25 cm and a length of 50 cm, giving a volume of 100 litres. The salt density was taken as 1.94 g/cc;  this was adjusted to compensate for the added uranium
and thorium assuming
that the masses and volumes of the components add. Exact figures depend on the chemistry of the solution and will require some investigation.
The cylinder of salt was enclosed in a cylinder of graphite 4m long and 2m in radius.

\begin{figure}[H]
\centerline{\includegraphics[width=10 cm]{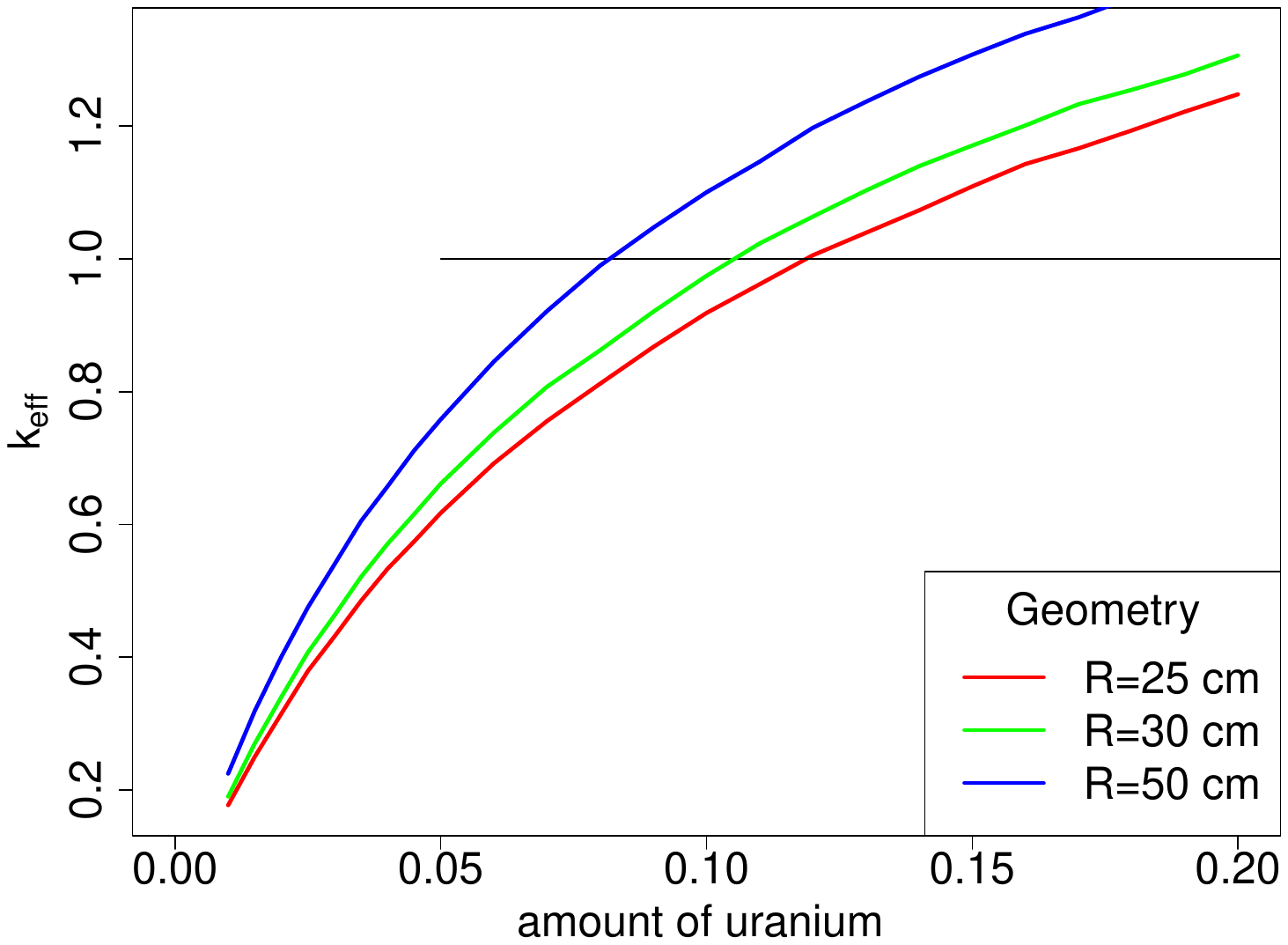}}
\caption{\label{fig:critical} Criticality $k$ as a function of \U\ concentration for various geometries. (Units are moles of uranium per mole of FLiBe.)}
\end{figure}

Results are shown in Figure~\ref{fig:critical}. The desired criticality is slightly below 1.0, and this
is achieved with concentrations of around 0.1 \U\ atoms per for each FLiBe molecule (which has a molecular weight of 99). 
100 litres of FLiBe weighs about 200 kg. For a 10\% \U\ solution that implies $200 \times 0.1 \times {233 \over 99}=47$ kg of \U. 
Larger radii require lower concentrations, due to smaller loss of neutrons to the outside, but the larger volume increases the mass of \U\ that would be required.

\begin{figure}[H]
\centerline{\includegraphics[width=9 cm]{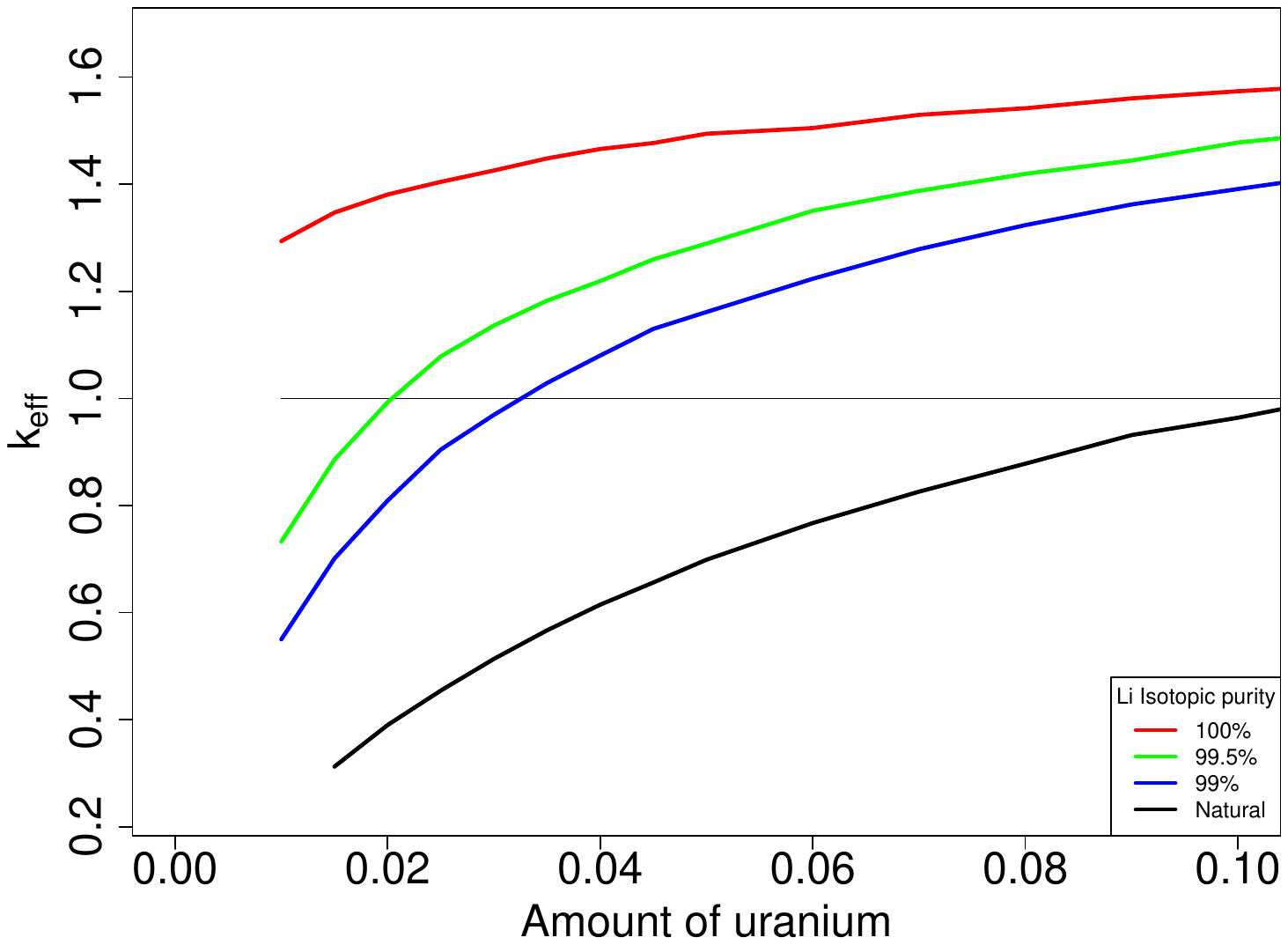}}
\caption{\label{fig:critical2} Criticality $k$ as a function of \U\ concentration for various lithium purities}
\end{figure}

The main loss of neutrons is due to absorption on ${}^6$Li, which has a very large neutron cross section.
Natural lithium is 92.5\% 
${}^7$Li and 7.5\%
${}^6$Li. The neutronics can be greatly improved by using isotopically purified lithium. Figure~\ref{fig:critical2} shows the criticalities as a function of \U\ concentration for various purities.
If a satisfactory criticality can be achieved with around 1\% concentration of \U, which the figure indicates is possible with a purity at the per mille level, then
the mass of \U\ required is reduced to a few kg which should be achievable.

\subsection{Criticality when other elements are added}\label{Criticality}

We took the salt with 99.5\% isotopically pure ${}^7$Li and 2\% uranium added as a baseline, and looked at the effect on the criticality of the addition of the tantalum, lead  and the thorium.

\begin{figure}[H]
\centerline{\includegraphics[width=8cm]{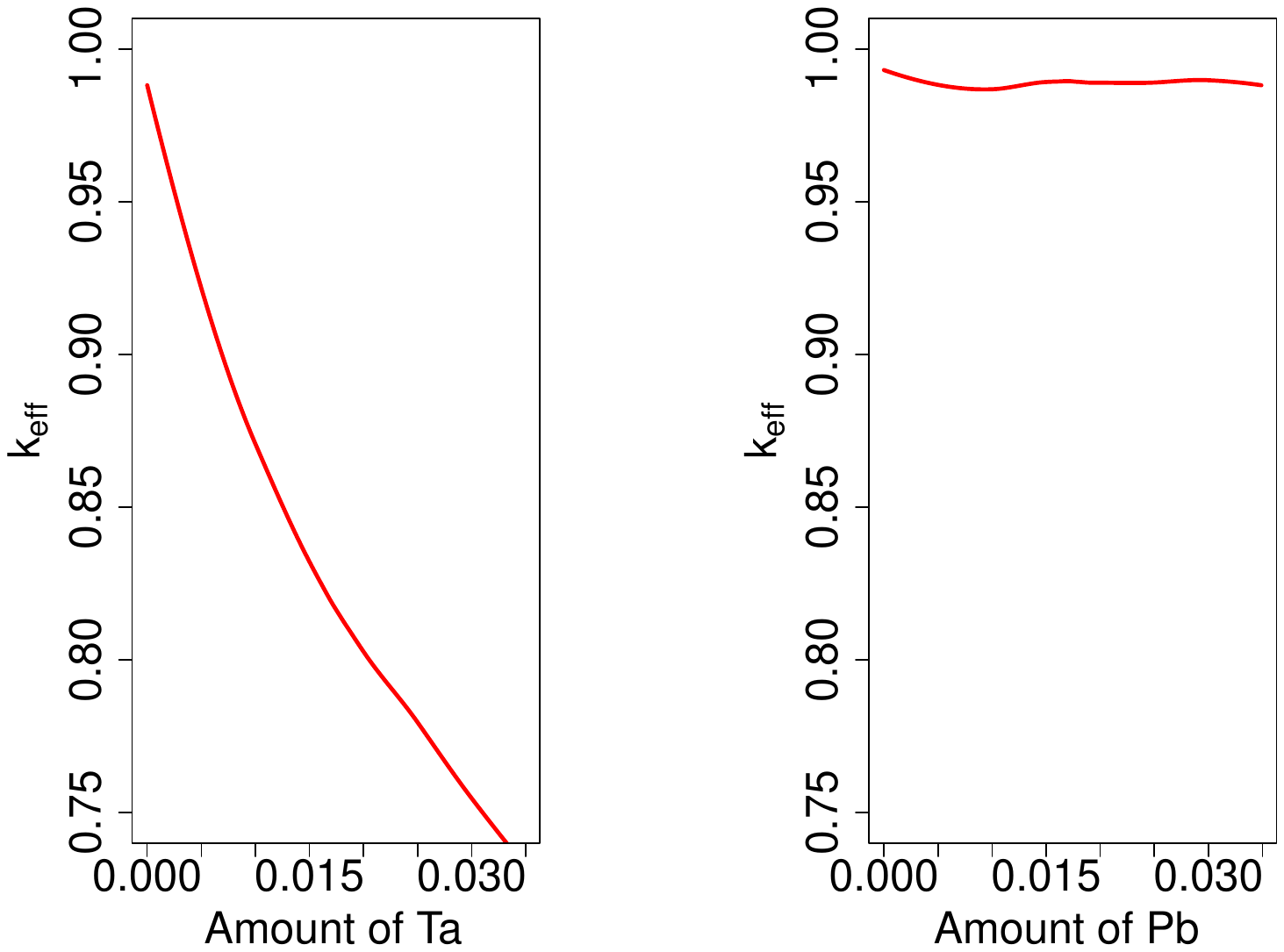}}
\caption{\label{fig:critical3} Criticality $k$ as a function of tantalum and lead concentrations. (Statistical fluctuations have been smoothed by the R loess algorithm) }
\end{figure}

Figure~\ref{fig:critical3} shows that adding tantalum reduces the criticality substantially, and this will have a large effect on any optimised design.  If lead (taken as pure \Pb) is used instead, this has very little effect on the criticality, as is also shown; this isotope is essentially transparent for neutrons. 
Lead will presumably generate photons in showers in much the same way that tantalum does should this be necessary.
\begin{figure}[H]
\centerline{\includegraphics[width=8 cm]{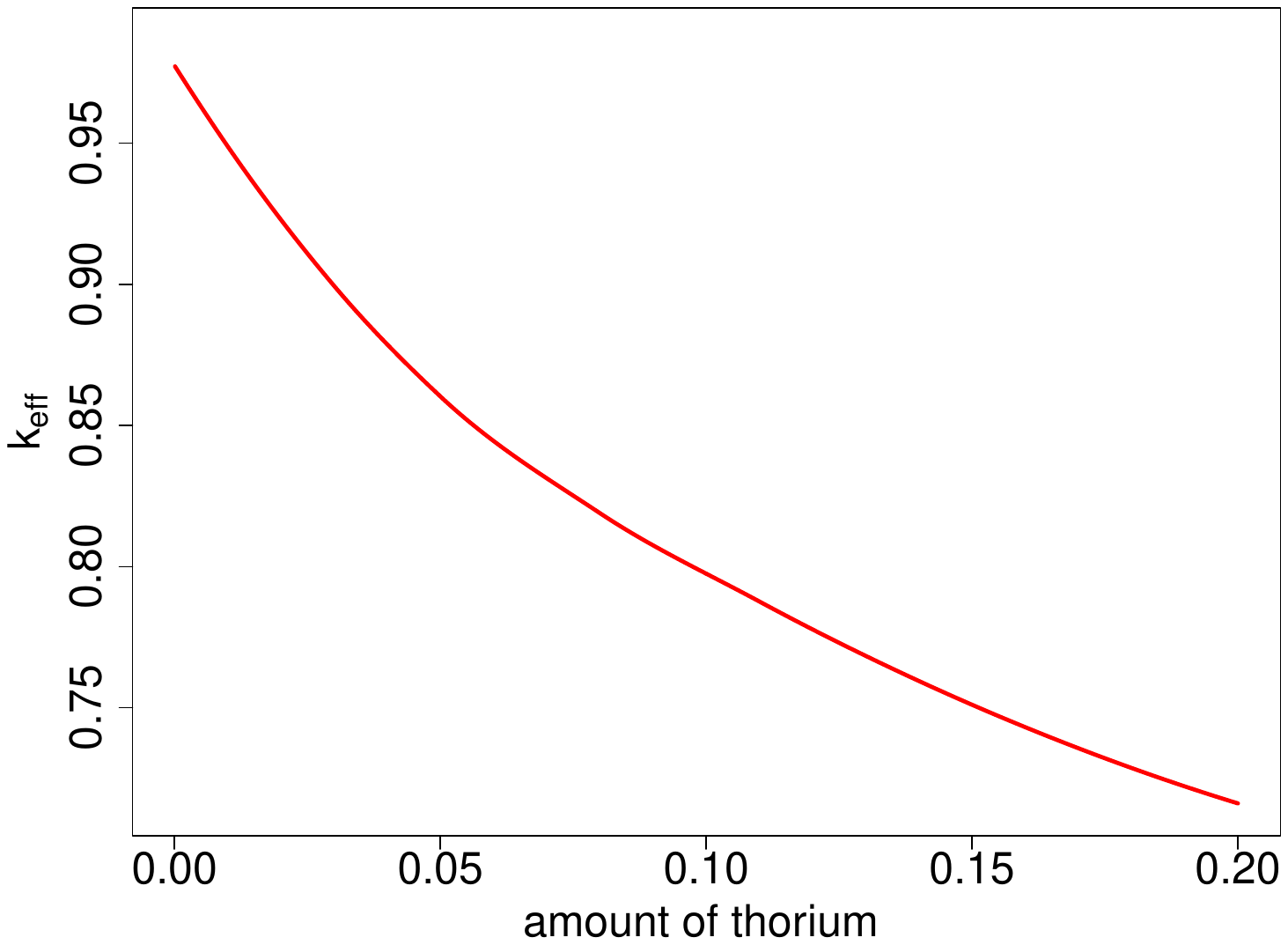}}
\caption{\label{fig:critical4} Criticality $k$ as a function of thorium concentration. (Smoothed as above.)}
\end{figure}

Figure~\ref{fig:critical4} shows that adding thorium also decreases the criticality, as one would expect  This means that the addition of thorium for breeding lowers the criticality - inevitably, as neutrons used for breeding are not used for fission.

\section{Neutron Production}
The number of neutrons per electron is recorded from Geant4. The photonuclear process is modelled in Geant4 using the G4EmExtraPhysics Physics list which includes G4PhotoNuclearCrossSection. In the
new models of Geant4 (version 11 and above), G4GammaNuclearXS is incuded in the form of G4PARTICLEXS-4.0 dataset which is based on IAEA evaluated photonuclear data library IAEA/PD-2019~\cite{Photonuclear}.

\begin{figure}[H]
\centerline{\includegraphics[width=9 cm]{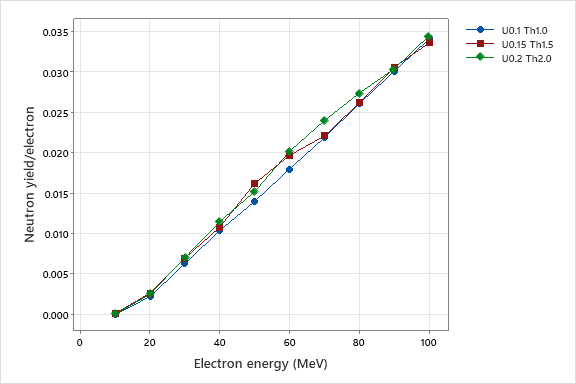}}
\caption{\label{fig:ConcThU} Number of neutrons/electron as a function of electron energy for different concentrations of uranium and thorium}
\end{figure}

Figure~\ref{fig:ConcThU} shows the neutron production in the salt with uranium and thorium for different concentrations. Only a little variation is observed in neutron numbers by changing the concentrations of uranium and thorium.

\begin{figure}[H]
\centerline{\includegraphics[width=10 cm]{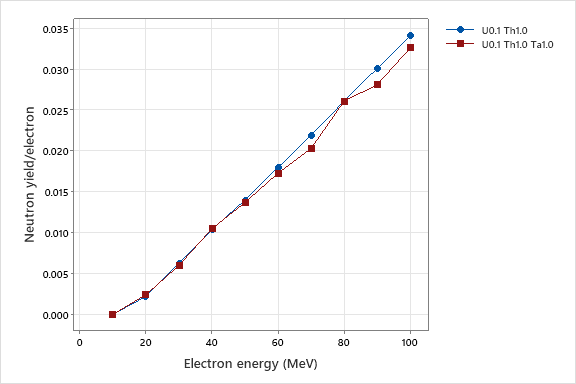}}
\caption{\label{fig:nTa} Effect of adding tantalum on neutron production}
\end{figure}

\begin{figure}[H]
\centerline{\includegraphics[width=10 cm]{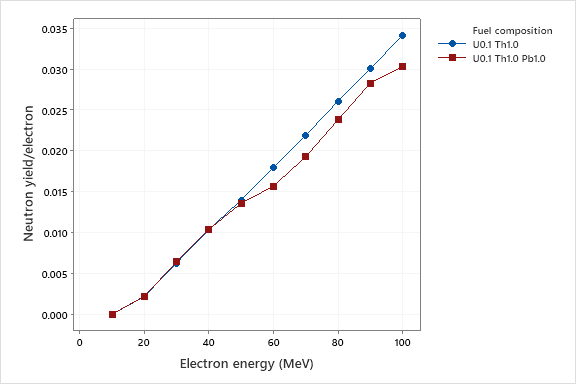}}
\caption{\label{fig:nPb} Effect of adding lead on neutron production}
\end{figure}

To observe the effect of adding tantalum and lead on photon production, these elements are added in the mixture with the concentration: U - 0.1 mole, Th - 1 mole, Pb/Ta - 1 mole. Figure~\ref{fig:nTa} and Figure~\ref{fig:nPb} show the effect of adding tantalum and lead in the mixture respectively. The figures show that although significant increase in neutron yield is observed by increasing the beam energy, for both tantalum and lead, there is no benefit of adding tantalum and lead on neutron production and after 50 MeV the production rather decreases with the addition of tantalum or lead.

This validates our earlier assertion that nothing is to be gained by adding high-Z material in an attempt to boost photon production.
At the optimum electron energy of 60 MeV about 0.02 neutrons are produced for each incident electron.

\begin{figure}[H]
\centerline{\includegraphics[width=13 cm]{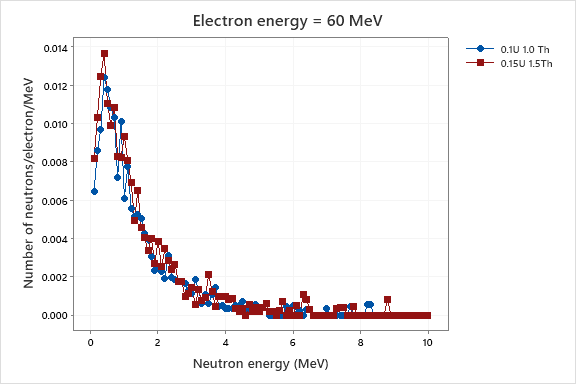}}
\caption{\label{fig:Spectra60MeV} Neutron energy spectra at 60 MeV electron energy 
}
\end{figure}

The energy distribution of photoneutrons is shown in Figure~\ref{fig:Spectra60MeV} for two fuel compositions (U - 0.1 mole/Th - 1.0 mole and U - 0.15 mole/Th - 1.5 mole). We get a peak corresponding to 0.014 neutrons/MeV at around 1 MeV, which will be taken as the typical photoneutron energy.

\section{Breeding}

\begin{figure}[ht]
\centerline{
\includegraphics[width=16 cm]{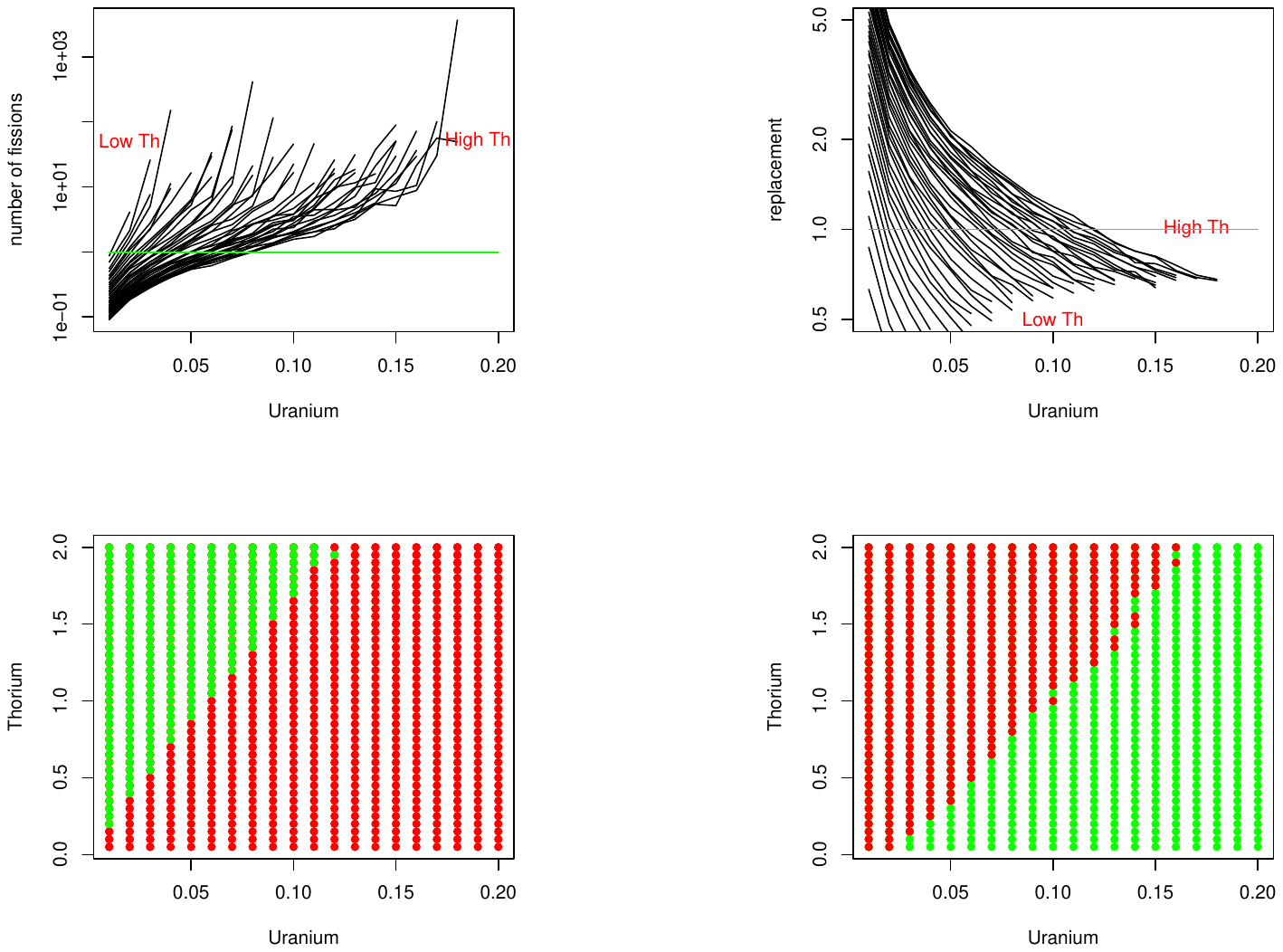}
}
\caption{\label{fig:breeding3} Criticality 
and replenishment rates for different concentrations of uranium and thorium.
The top left shows the number of fissions per initial neutron as a function of uranium concentration, for different thorium concentrations (higher concentrations give fewer neutrons).
The top right shows the number of $(n,\gamma)$ conversions on thorium, leading to replacement of the fertile \U\ (higher concentrations give more replacement). 
The bottom left shows the region of uranium and thorium concentration where the 
replacement exceeds 1, the bottom right the region where the number of fissions per neutron exceeds twenty 
}
\end{figure}

As the \U\  is consumed, it will be replaced using the fertile-to-fissile conversion.

$$ {}^{232}{\rm Th} \qquad  { {(n,\gamma)} \atop \longrightarrow } \qquad 
{}^{233}{\rm Th} \qquad { \beta \atop \longrightarrow} \qquad   {}^{233}{\rm Pa} \qquad { \beta \atop \longrightarrow } \qquad {}^{233}{\rm U}$$

In use, the replacement rate must equal the rate at which fuel is consumed.  As the cross sections for the $(n,\gamma)$ process are roughly a factor of 10 below the relevant fission cross sections,
this will require a thorium concentration about ten times that of the uranium.

In this simple study we ignore the effect of fission products and also the intermediate protactinium, which 
can also absorb neutrons leading to undesired isotopes. 
Its half-life of 27 days is longer than the equivalent in the uranium cycle, and
this may be long enough for it to affect the neutronics,
reducing the breeding rate.
If there is a problem then
a process can be implemented to continually remove fission products and 
separate the protactinium, so that it can decay to \U\ in an environment with no neutrons present.  
Safeguards will be necessary to ensure that the fissile \U\ is not diverted for military purposes.
One must also be aware of the `protactinium problem': if the reactor is switched off then
no fuel is being consumed, but as the protactinium decays it will increase the \U\ concentration and may drive the reactor 
into and above criticality.

The addition of thorium reduces the reactivity -- for a breeder reactor the 2.5 average neutrons per fission
leaves only 0.5 neutrons for losses so the design is severely constrained.
This has been studied using MCNP with `tallies' for $(n,\gamma)$ reactions on \Th\ and for fission
on \U, with a small contribution from \Th. 
Results from the fission rates are not identical to the criticality considered in
the previous section, as the initial neutron energy (taken as 1 MeV: in a full study this should be replaced by the spectrum from Figure~\ref{fig:Spectra60MeV}) is not the same.
Figure~\ref{fig:breeding3} shows the number of fissions per inital neutron 
as a function of uranium concentration (the axis is the number of \U\ nuclei per \FLiBe `molecule')
for \Th\ concentrations from 0.01 to 2.0, in the same units. 
The top right plot shows the replacement rate divided by the number of fissions, which
ideally should equal 1.0.   For a given \U \ concentration, the replacement rate increases with the thorium concentration, but the fission rate is reduced, as the thorium absorbs more neutrons. For stable operation
we require a criticality of 0.98 and replacement rate of 1.0, shown by the green lines. It can be seen
that for a given uranium fraction,  criticality requires low thorium and replacement requires high thorium.

The lower plots show the concentrations of \U\ and \Th\ for which the replacement rate is
adequate (left) and the fission rate adequate (right).
Regrettably there does not seem to be a region where both can be satisfied. 
To remedy this we will have to
\begin{itemize}
    \item Relax the breeding requirement, so that the thorium lengthens the fuel lifetime, but not indefinitely.
    \item Relax the reactivity requirement, if efficiency savings can be found from somewhere else
    \item Increase the size of the reactor to reduce geometric losses, or improve neutron reflection
    \item Devise some way of increasing the neutron efficiency by shaping the spectrum.
\end{itemize}

\begin{figure}[ht]
\centerline{
\includegraphics[width=8 cm]{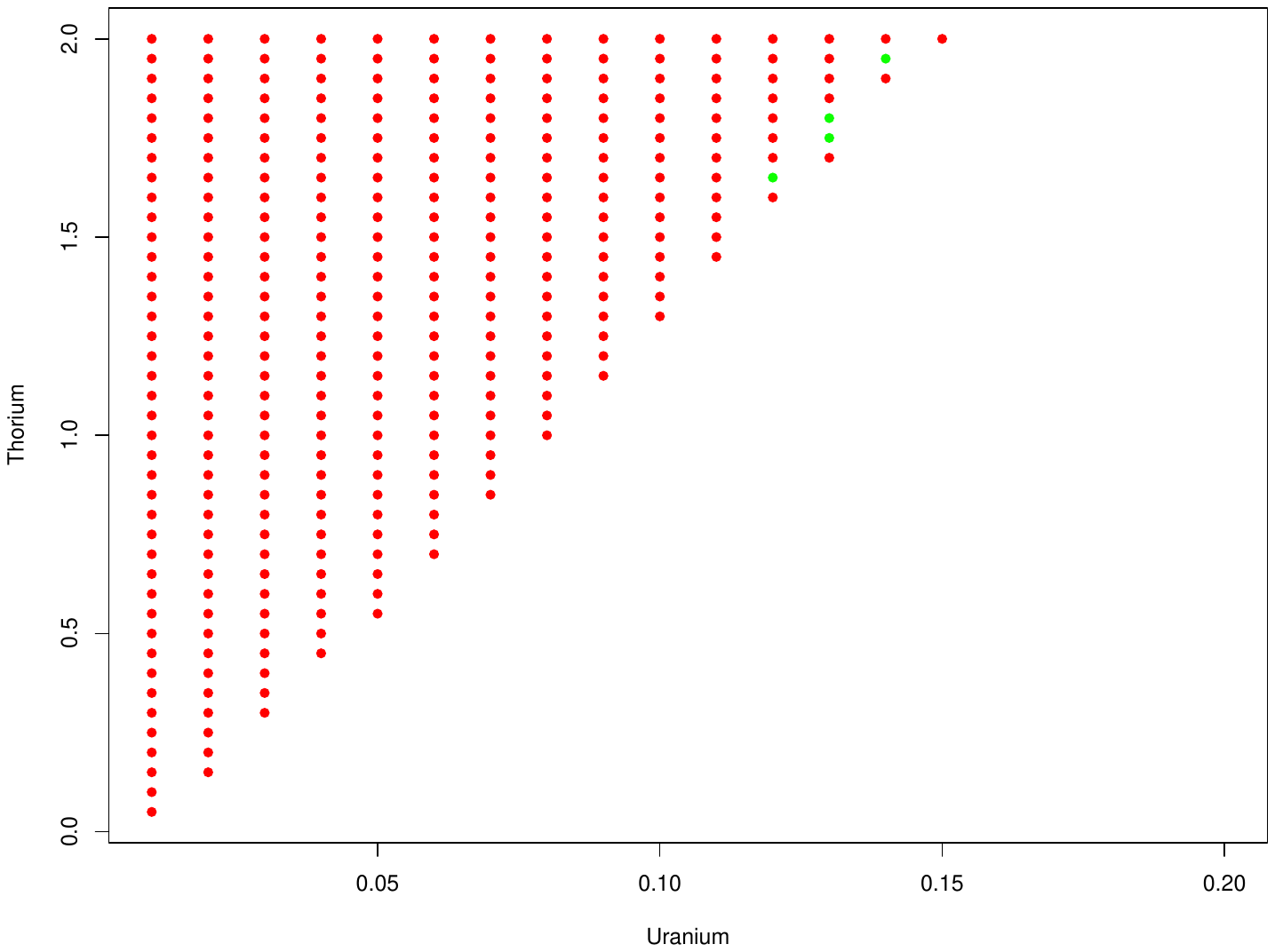}
\includegraphics[width=8 cm]{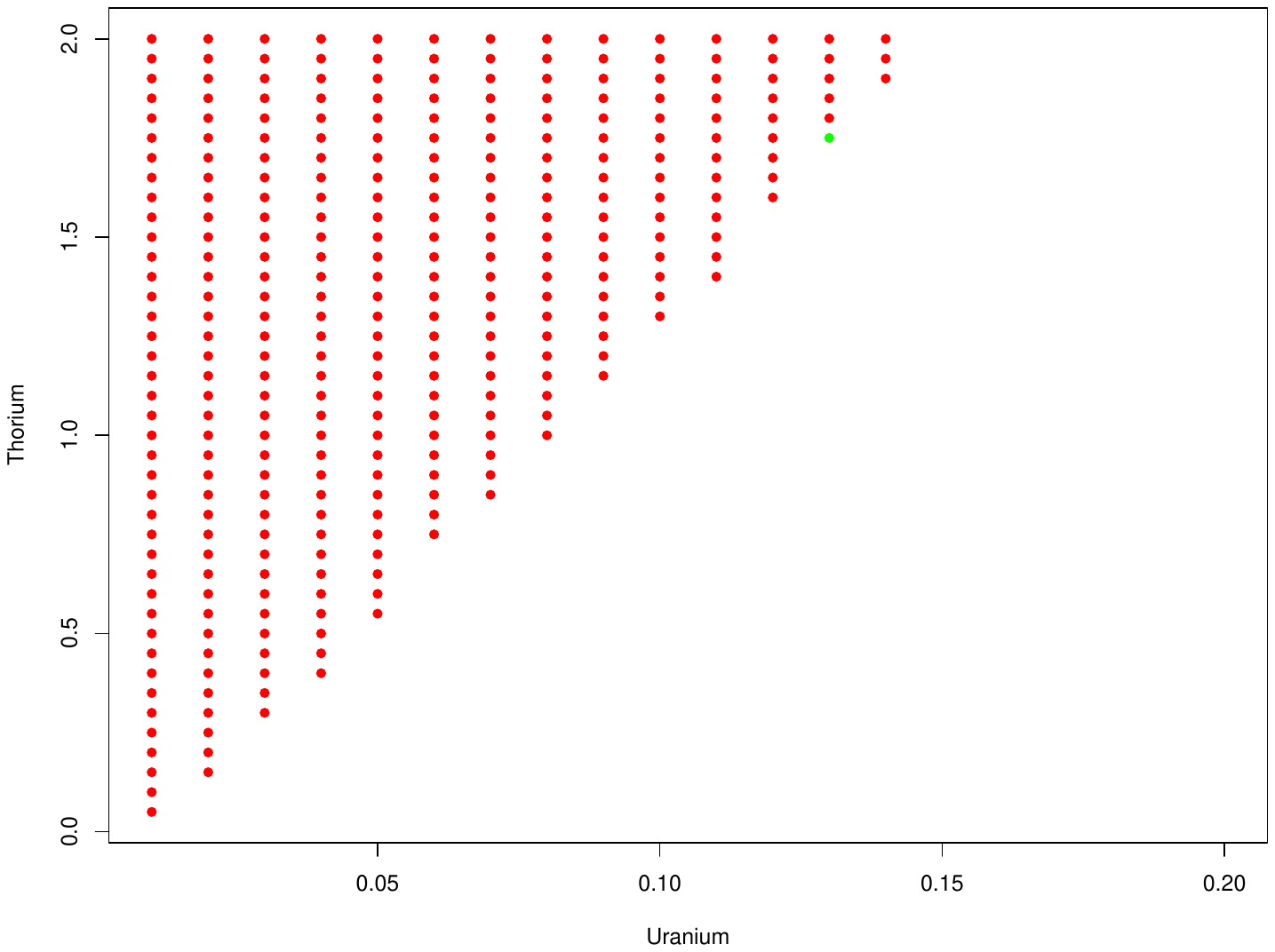}
}
\caption{\label{fig:breeding4} 
Combination of the replenishment and criticality requirements for a larger reactor. 
Outcomes with a criticality greater than 1 are not shown. The second plot has 0.1 atoms of \Pb  \ added.
}
\end{figure}

As an example, we tried doubling the linear size of the molten salt cylinder (which would give 8 times the volume). Results are shown in Figure~\ref{fig:breeding4}.  Combinations shown in red have either too low a replacement rate or too low a neutron multiplication (or both). 
Satisfactory outcomes are shown in green, and with the reduced geometric loss we do now have
some possibilities.   If 0.1 atoms of \Pb\ per FLiBE are added to boost the photoneutrons, there is still a small region of viability, as shown in the second plot. 
This needs  to be explored further.

\section{Summary}


We consider the design of a micro modular sub-critical molten salt breeder reactor, driven by an electron accelerator.
The seed fissile material is \U, and this is continually replaced during operation by breeding from the thorium
present.

No purpose is served by adding tantalum (or lead) beads. To obtain near-criticality, $k\approx 0.98$, and complete fuel replacement
requires concentrations in the region of 1 \Th\ atom per FLiBe, and 0.1 \U\ atoms.



Criticalities between 0.9 and 1.0 can be achieved
despite the addition of 
thorium to the uranium-loaded FLiBe salt. However the lithium in the mixture will have to be isotopically purified, to remove the ${}^6$Li which has a high neutron absorption cross section.
The exact criticalty will need to be decided carefully: economic viability
demands a value close to 1.0, whereas safety requires it to be below this by a margin wider than any possible fluctuation. The negative temperature coefficient of a molten salt reactor will be some help here.

For a viable power source the energy produced has to exceed the energy 
put in:

\begin{equation}
    \eta_1 {E_{f} \over 1-k} r \nu > {E_e\over \eta_2}
\label{eq:viability}
\end{equation}
where $\eta_1$ is the efficiency for thermal to electrical power conversion, $E_e$ is the electron energy, $\eta_2$ is the efficiency of the accelerator, $E_f$ is the energy from fission, $k$ is the criticality, $\nu$ is the number of photoneutrons produced by an electron and $r$ is the fraction of these which cause fission.  If we take $E_e=50 MeV$ and (optimistically) 
$\eta_1=0.5, \eta_2=1, r=0.4, E_f=200 MeV, k=0.98$ this gives the requirement $\nu>0.025$: each electron has to give 0.025 photoneutrons.  
 At 60 MeV, 0.02 photo neutrons per electron are obtained which is  slightly below the 0.03 needed for viability, given by Equation~\ref{eq:viability}.
 The current design is thus on the edge of viability.
 This could be relaxed by increasing the criticality, although this has implications for safety.

These calculations are heavily dependent on details of the cross sections for many isotopes at many energies. 
To produce a design for actual construction, 
a programme of measurements will need to be carried out using electron beams on FliBe, with various additives, and 
the number and energies of photoneutrons produced measured experimentally. Such an experiment is fundamental to taking this design further.


\begin{thebibliography}{9}

\def \etal {{\em et al.}}
\bibitem{SMR}
Small Modular Reactors. https://www.iaea.org/topics/small-modular-reactors [Accessed: 19-03-2024]

\bibitem{ARIS}
International Atomic Energy Agency (2020). Advances in Small Modular Reactor Technolo-gy Developments: A Supplement to the IAEA Advanced Reactors Information System (ARIS). IAEA.

\bibitem{Nifenecker} 
H. Nifenecker, S. David, J.M. Loiseaux, O. Meplan, Basics of accelerator driven subcritical reactors, Nuclear Instruments and Methods in Physics Research Section A: Accelerators, Spectrometers, Detectors and Associated Equipment, 463 3 (2001)

\bibitem{Asiya}
A. Rummana, Spallation Neutron Source for an Accelerator Driven Subcritical Reactor. Doctoral dissertation, University of Huddersfield (2019).


\bibitem{AccApp} 
R J Barlow, Thorium Fuelled Reactors: Do they need an accelerator?,
Proc. 12th InternationalTopical Meeting on Nuclear Applications of Accelerators (AccApp15),
American Nuclear Society (2017)

\bibitem{Myrrha} 
A Rummana, R J Barlow, S M Saad, Calculations of neutron fluxes and isotope conversion rates in a thorium-fuelled MYRRHA reactor, using GEANT4 and MCNPX. Nuclear Engineering and Design. 388:111629 (2022).

\bibitem{Zhao}
Zhao, X. C., Cui, D. Y., Cai, X. Z., and Chen, J. G. (2018). Analysis of Th-U breeding capability for an accelerator-driven subcritical molten salt reactor. Nuclear Science and Techniques, 29(8), 121.

\bibitem{previous}
G. Myneni, Virginia ADS Consortium - Advanced \U\ - \Th\ Breeder Burner Subcritical Micro Reactors (ASMR). 
{\footnotesize 
\begin{verbatim} 
https://msrworkshop.ornl.gov/wp-content/uploads/2020/11/35_Myneni_MSRW_ORNL_20201.pdf
\end{verbatim}
}
[Accessed:17.03.2024]

\bibitem{Abalin}
S. S. Abalin et al., Conception of electron beam-driven subcritical molten salt ultimate safety reactor, AIP publishing, U.S.A. (1995).
\bibitem{Liu}
Y. Liu, A study on the feasibility of electron-based accelerator driven systems for nuclear waste transmutation, Ph.D. Thesis, North Carolina State University (2006).
\bibitem{Feizl}
H Feizi and A.H. Ranjbar,Developing an Accelerator Driven System (ADS) based on electron accelerators and heavy water, J. Inst 11 P02004 (2016)

\bibitem{Zelinsky}
A Y Zelinsky et al, NSC KIPT Neutron Source On The Base Of Subcritical Assembly Driven With Electron Linear Accelerator, IPAC13.

\bibitem{DAFNE1}
L. Quintieri \etal,
{\em Photoneutron Source by High Energy Electrons on High Z Target: Comparison Between Monte Carlo Codes and Experimental Data}, Transaction of Fusion Science and Technology {\bf 61} 313 (2012)

\bibitem{DAFNE2}
L. Quintieri \etal,
A Photoneutron source at the Da$\Phi$ne Beam Test Facility of the INFN National Laboratories in Frascati: design and first experimental results
Physics Procedia {\bf 26}249 (2012)

\bibitem{Gillespie}
A K Gillespie, C Lin and R V Duncan,
``Photoneutron Yield for an electron beam on tantalum and Erbium deuteride",
arXiv :2308.02629 (2023)

\bibitem{IMSBR} 
I. V. Dulera. ``Indian Molten Salt Breeder Reactor-challenges and developments for thermal techniques.'' Proceedings of the twenty second DAE-BRNS symposium on thermal analysis-thermal techniques for advanced materials: book of abstracts. 2020.

\bibitem{BARC} 
BARC activities for Indian Nuclear Power Program. https://www.barc.gov.in/randd/ [Ac-cessed: 20-08-2023]

\bibitem{EGS}
H Hirayama et al.,``The EGS5 code system'', SLAC Report SLAC-R-730 (2015)

\bibitem{Geant4}
S. Agostinelli \etal, GEANT4-a simulation toolkit, Nucl. Instrum. \& Meth. A {\bf 506},
    p250
    (2003)
\bibitem{AllisonNucl}
J. Allison \etal, Recent Developments in Geant4, Nucl. Instrum. Meth. A {\bf 835} 186-225 (2016) 
\bibitem{AllisonIEEE}
J. Allison \etal, Geant4 Developments and Applications, IEEE Trans. Nucl. Sci. {\bf 53} 270-278 (2006)

\bibitem{xsect}  
C W  Arnold et al, {\em A new absolute total cross-section for photodisintegration of beryllium-9}, arXiv:1112.1148v2 [nucl-ex]

\bibitem{ENDF}
Evaluated Nuclear Data File (ENDF). https://www-nds.iaea.org/exfor/endf.htm \hfil  [Accessed: 09.09.2023]
 
\bibitem{MCNP}
X-5 Monte Carlo team, ``MCNP - Version 5, ol 1: Overview and theory", LA-UR-03-1987 (2003)

\bibitem{Photonuclear} 
B. Kutsenko. New Geant4 photonuclear cross-section model. No. CERN-STUDENTS-Note-2021-034. 2021.

\end{thebibliography}
\end{document}